\documentclass[12pt]{article}

\newcommand{\be}{\begin{equation}}
\newcommand{\ee}{\end{equation}}
\newcommand{\ba}{\begin{eqnarray}}
\newcommand{\ea}{\end{eqnarray}}
\newcommand{\no}{\noindent}
\newcommand{\n}{\label}

\begin{document}

\title{Power-law expansion in $k$-essence cosmology}

\author{Luis P. Chimento  $^{1,\,\dag}$ and Alexander Feinstein $^2$\\
{$ ^1$ \small Dpto. de F\'{\i}sica,
Facultad de Ciencias Exactas y Naturales, }\\
{\small  Universidad de Buenos Aires, Ciudad  Universitaria}\\
{\small Pabell\'on  I, 1428 Buenos Aires, Argentina.}\\
{$ ^2$ \small Dpto. de F\'{\i}sica Te\'orica,}\\
{\small  Universidad del Pa\'{\i}s
  Vasco,}\\
{\small   Apdo. 644, 48080, Bilbao, Spain}.}

\maketitle

\begin{abstract}

We study spatially flat isotropic universes driven by $k$-essence. It is shown
that Friedmann and $k$-field equations may be analytically integrated for
arbitrary $k$-field potentials during evolution with a constant baryotropic
index. It follows that there is an infinite number of dynamically different
$k$-theories with equivalent kinematics of the gravitational field. We show
that there is a large ``window" of stable solutions, and that the dust-like
behaviour separates stable from unstable expansion. Restricting to the family
of power law solutions, it is argued  that the linear scalar field model, with
constant function $F$, is isomorphic to a model with divergent speed of sound
and this makes them less suitable  for  cosmological modeling than the
non-linear $k$-field solutions we find in this paper.

\end{abstract}

\vskip 1cm

\noindent\dag{Fellow of the Consejo Nacional de Investigaciones
Cient\'{\i}ficas y T\'ecnicas.}

\newpage

%============================================================
\section{introduction}
%==================================================================

To address the outstanding theoretical challenges of modern cosmology,
especially the so-called coincidence problem, which questions as why is it
exactly now, the universe driven by some sort of dark energy is accelerating,
several authors have introduced and studied the so-called $k$-essence models
\cite{armendariz1,armendariz2,chiba,malquarti}.

Originally, the $k$-essence, or $k$-inflation, was introduced in
\cite{armendariz3} in order to bridge phenomenologically the string
theories with inflation (see Ref. \cite{quevedo} for a recent review). The main ingredient of the $k$-essence is a scalar
field, with non standard higher order kinetic  terms. Interestingly enough, and contrary to what one could have expected, these non-standard terms do not necessarily lead to
acausal propagation of the $k$-field \cite{garriga}. Studying inflationary patterns with the $k$-fields the authors of
\cite{armendariz3}  were able to show that $k$-field may drive an
accelerated expansion of the universe starting from a generic initial
conditions without an assistance of the usual potential terms.

In a different development \cite{armendariz1,armendariz2}, the $k$-essence was
proposed as a dynamical solution to the coincidence problem. The basic idea of
\cite{armendariz1,armendariz2} is that $k$-essence could play a role of a dynamical
attractor at the onset of matter domination period introducing cosmic acceleration at
present time. Further study of $k$-essence was performed recently in
\cite{malquarti}. It was argued that in certain dynamical regimes the
$k$-essence is equivalent to quintessence and it may prove difficult to
distinguish between the two fields. In this paper we  make a step further and  show that the dynamically different $k$-theories can produce kinematically equivalent cosmological models.

The construction of cosmological models with tracker-like, or the attractor behaviour \cite{trac}, where the
$k$-essence either mimics the equation of state of the matter-radiation
component, or drives towards acceleration, is relied heavily on the existence
of $k$-essence solutions which, re-written in terms of energy density and
pressure, represent, hydrodynamically, fluids with a constant baryotropic index (BI).
These, in turn, give rise to a power-law behaviour of the scale factor when the
underlying geometry is that of a spatially flat isotropic universe.

In the $k$-essence models studied earlier
\cite{armendariz1,armendariz2,chiba,malquarti} one would usually consider
solutions where, during  the $k$-field driven expansion with the constant BI,
two things happen: i) The scalar field evolves linearly with time and ii) The
$k$-field potential is an inverse square in terms of the $k$-field. The
property ii) follows directly from i). The assumption i), triggered probably
by the simplicity of finding solutions in the case of linear $k$-field,
although permits to consider different $k$-theories, is too restrictive with
the form of the k-potential and the evolution of the field itself.

The main purpose of this paper is to show that in the case of the
$k$-essence, one may find solutions with arbitrary potentials and non-linear
scalar fields, but still have a constant BI.  For the solutions we find, one
can have a fixed evolution of the geometry, yet incredibly rich repertoire of
scalar field behaviour and its $k$-potential. This kind of a degeneracy is
quite  problematic for the model building. Nevertheless, there seems to be a
way to reduce this degeneration. For solutions with constant BI we find that
the linear $k$-field model is isomorphic to a divergent sound speed model. In
fact, the former presents a superior type of degeneracy as compared to the
solutions obtained from the non-linear $k$-field model, therefore fully
justifying our quest for a different type of $k$-field solutions. The
isomorphism between the linear $k$-field model and the divergent sound speed
model looks especially interesting in the light of the results of the recent
publication \cite{malquarti2} where the behaviour of the  solutions near
divergent sound speed was thoroughly investigated.

%==================================================================
\section{The general framework}
%==================================================================

We start with a general Lagrangian

\be
\n{l}
{\cal L}=-V(\phi)\,F(x),   \qquad   x=g^{\mu\nu}\phi_\mu\phi_\nu,
\ee

\no where $\phi$ is the scalar field and $\phi_\mu=\partial\phi/\partial
x^\mu$, and  do not impose any conditions neither on $V$, nor $F$ at this stage.
One may easily figure out the energy-momentum tensor for (\ref{l}):

\be
\n{t}
T_{\mu\nu}=V(\phi)\left[2F_x\phi_\mu\phi_\nu-g_{\mu\nu}F\right],
\qquad F_x=\frac{d\,F}{d\,x}.
\ee

\no Identifying (\ref{t}) with the energy-momentum tensor of a perfect fluid
we have

\be
\n{ro}
\rho_\phi=V(\phi)[F-2xF_x],  \qquad   p_\phi={\cal L}=-V(\phi)F.
\ee

\no As usual in this setting we  assume a spatially flat homogeneous and isotropic
spacetime with line element

\be
\n{rw}
ds^{2}=-dt^{2}+a^{2}(t)\left[dx^{2}+dy^{2}+dz^{2}\right],
\ee

\no where $a(t)$ is the scale factor and the expansion rate is defined as $H=\dot a/a$.
The Einstein field equations then reduce to

\be
\n{00}
3H^2=\rho_\phi,      \qquad  -2\dot H=\rho_\phi+p_\phi,
\ee

\no and the conservation equation reads

\be
\n{con}
\dot\rho_\phi+3H(\rho_\phi+p_\phi)=0.
\ee

\no The field equation for the $\phi$ field may be either obtained by
substituting expressions (\ref{ro})  into the conservation
equation (\ref{con}), or varying directly the Lagrangian (\ref{l}). Doing so,
we get:

\be
\n{kg}
[F_x+2xF_{xx}]\ddot\phi+3HF_x\dot\phi+\frac{V'}{2V}[F-2xF_x]=0,
\ee

\no where $V'=dV/V\phi$. On the other hand, assuming a ``formal" equation of
state of the form $p_\phi=(\gamma-1)\rho_\phi$ for the k-essence  and using
Eqs. (\ref{ro}), (\ref{00}) we obtain the BI $\gamma$

\be
\n{ga}
\gamma=-\frac{2\dot H}{3H^2}=-\frac{2xF_x}{F-2xF_x}.
\ee

\no We now assume that the BI is a constant. This
kinematically leads to a power-law scale factor $a=a_0t^{2/3\gamma}$.

The first question we ask is, how stable are the solutions with the constant
BI $\gamma=\gamma_0$?.  To answer this question, we allow $\gamma$ to vary
with time. Differentiating the equation of state and using the conservation
equation we find

\be
\n{p.}
\dot\gamma=3H\gamma(\gamma-1)+\frac{\dot p_\phi}{\rho_\phi},
\ee

\no which together with (\ref{ro}) and (\ref{ga}) lead to

\be
\n{ga.}
\dot\gamma+\left[3H\gamma+\frac{\dot V}{V}+
\frac{\dot F}{F}\right](1-\gamma)=0.
\ee

\no We further check as to whether $\gamma=\gamma_0$ are  solutions to this
equation at all. Obviously, there are two different ways for this  to happen: either $\gamma_0=1$, or generically,  the following stationary condition  holds

\be
\n{at}
\frac{\dot V}{V}+\frac{\dot F}{F}=-3H\gamma_0,
\ee

When the stationarity condition (\ref{at}) holds, the
potential $V$ and the function F are related by:

\be
\n{vf}
VF=\frac{4(1-\gamma_0)a_0^{3\gamma_0}}{3\gamma_0^2a^{3\gamma_0}}.
\ee

\no Here, we have integrated (\ref{at}) and inserted the solution into the
Einstein equation (\ref{00}) to fix the integration constant. For a positive
potential $V$, the constrain (\ref{vf}) gives rise to two different theories
depending on whether  $\gamma_0<1$ or $\gamma_0>1$. In the case $\gamma_0<1$
we take the function $F$
to be positive, whereas in the case $\gamma_0>1$ we take it  negative. We denote these
as $F^+$ and $F^-$ respectively.

We now assume that the stationarity condition (\ref{at}) holds. So, the
equation (\ref{ga.}) reads

\be
\n{ga.f}
\dot\gamma+3H(\gamma-\gamma_0)(1-\gamma)=0.
\ee

\no Integrating, we find:

\be
\n{gg}
\gamma=\frac{\gamma_0+c\,a^{-3(1-\gamma_0)}}{1+c\,a^{-3(1-\gamma_0)}}.
\ee

\no Here $c$ is an integration constant. For the expanding universe and
$\gamma_0<1$ we see that the solutions of (\ref{ga.f}) have the asymptotic
limit $\gamma_0$. Therefore, the solutions with constant BI $\gamma_0$ are
attractors in the case $\gamma_0<1$. This attractor behaviour holds even for
superaccelerated universes \cite{negative} with $\gamma_0<0$.

The limit $\gamma_0\to 1$, should be considered apart, and the  solution of
the equation (\ref{ga.f}) is

\be
\n{g1}
\gamma=1-\frac{1}{c+\ln a^3},
\ee

\no where $c$ is an integration constant. Hence, for an expanding universe the
solution with $\gamma_0=1$ is stable  as well . The  $\gamma_0=1$ solutions
separate stable from unstable regions  in the phase space (for a positive
expansion rate) as can be easily seen from the equation (\ref{ga.f}), and
since $\gamma_0=1$ corresponds to dust, we conclude that the dust-like
solutions define the border line between stable and unstable behavior. It is
probably worthwhile to mention that the above stability analysis is simple and
direct  as compared to  the study performed directly in the field variables
using the solution $\phi \propto t$ as an input.

%==================================================================
\section{Power-law solutions}
%==================================================================

As from now we stick to the solutions with the constant BI $\gamma$. It
follows then that the Einstein and the field equations (\ref{00}),
(\ref{kg}) have {\it two} different classes of solutions:

\vskip .5cm

\no 1) The solutions with constant $x=x_0=-\dot\phi^2$.

\vskip .5cm

\no In this case for $\gamma=const\ne 0$ we have $a=a_0t^{2/3\gamma}$, the
first term of the l.h.s. of the Eq. (\ref{kg}) vanishes, and the consistent
solution of Eqs. (\ref{00}), (\ref{kg}) becomes $\phi=\pm\sqrt{-x_0}\,t$ and

\be
\n{V}
V=-\frac{4x_0}{3\gamma^2[F-2xF_x]}\frac{1}{\phi^2},
\ee

\no with  an arbitrary $F$ evaluated at $x=x_0$. We will not discuss  these
solutions further, since these were thoroughly investigated and exploited in model
building in \cite{armendariz1, armendariz2}. The particular case with $x_0=0$
 ($\phi=\phi_0$) must be solved apart and gives a de Sitter
solution $a=a_0{\mbox e}^{\sqrt{VF/3}\,t}$ for arbitrary $F$ evaluated at
$x=0$ and constant potential $V$.

\vskip .5cm

\no 2) Solutions with $x\ne const$.

\vskip .5cm

\no In this case the conservation equation (\ref{con}) can be readily integrated
to  find the first integral of the field equation (\ref{kg})

\be
\n{pi}
VF_\gamma=\frac{\rho_0}{a^{3\gamma}}.
\ee

\no Comparing this expression with  the constrain equation (\ref{00}) we are lead to the
relation (\ref{vf}) between the potential $V$ and the function $F$.
Hence, the integration constants $a_0$ and $\rho_0$ are left fixed to
$\rho_0=4(1-\gamma)a_0^{3\gamma}/3\gamma^2$.

We now look at (\ref{ga}) as a
differential equation for $F(x)$. Its immediate general solution is

\be
\n{fg}
F_\gamma(x)=c\,(-x)^{\frac{\gamma}{2(\gamma-1)}}.
\ee

\no Without any loss of generality, one may fix the integration
constant $c=\pm 1$. The  two corresponding families of solutions are then
 $F^+_\gamma$ and $F^-_\gamma$ respectively. Inserting the last equation
into (\ref{ro})  we get  two possibilities

\be
\n{rop+}
\rho^+_\phi=\frac{VF^+_\gamma}{1-\gamma}, \qquad  p^+_\phi=-VF^+_\gamma,
\qquad  \gamma<1,
\ee

\no and

\be
\n{rop-}
\rho^-_\phi=\frac{VF^-_\gamma}{1-\gamma}, \qquad  p^-_\phi=-VF^-_\gamma,
\qquad  \gamma>1,
\ee

\no where we have assumed that both the $k$-potential and the energy density are positive
definite.

Inserting (\ref{fg}) into (\ref{vf}) one gets a relation of the form
$t^{2}V\propto\dot\phi^{\gamma/1-\gamma}$. Finally, the general relations
connecting the field $\phi$ and the potential $V$ follows:

\be
\n{fi}
t^{\frac{2-\gamma}{\gamma}}=\frac{2-\gamma}{\gamma}
\left[\pm \frac{3\gamma^2}{4(1-\gamma)}\right]
\int V^{\frac{\gamma-1}{\gamma}}\,d\phi,  \qquad  \gamma\ne 2,
\ee

\be
\n{g2}
\ln t=\sqrt{3}\int \sqrt V d\phi,     \qquad     \gamma=2.
\ee

\no The $(+)$ branch in equation (\ref{fi}) corresponds to $\gamma<1$ while the
$(-)$ branch to $\gamma>1$. For linear $\phi$ the integral (\ref{fi}) is
not defined and this situation corresponds to the first class (i) of the
solutions. The relations (\ref{fi}) and (\ref{g2}) should be read as follows: given $V(\phi)$,
one may integrate and obtain $t=t(\phi)$, invert and find $\phi=\phi(t)$. Then
$F(x)$ is still given by (\ref{fg}). Note, that for a fixed $\gamma$ (fixed
power of the scale factor) one has different potentials and different field
evolutions, and consequently different $k$-theory. It looks
as the $k$-essence theories have a considerable amount of freedom in choosing
the theory, the potential and the scalar field behaviour, all describing the
same kinematics of the universe. This sounds somewhat ``fantastic" for these
are not just simple field redefinitions, and all the theories with the different $\phi$ and $V$ are dynamically different.

We now show how the power law  solutions with the linear scalar field and the
inverse square potential are related to the family of solutions with the
divergent velocity of sound. We do so by constructing a one-to-one mapping
between these solutions. We suggest that this might be the reason as to why
the solutions with the linear scalar field run into trouble as discussed in  a
recent paper by Malquarty etal \cite{malquarti2}.
 
 To do so we  introduce what we  call a ``divergent" $k$- essence Lagrangian
with the kinetic energy proportional to the velocity. For such a theory one
may take the function $F^\infty$ as

\be
F^\infty=c+b\sqrt{-x}.
\ee
 
It is easy to see that the above function  leads immediately to a divergent
sound velocity $C_s^2=-F^\infty_x/(F^\infty_x+2xF^\infty_{xx})$ and to an
inverse square potential by using the $k$-field equation (\ref{kg}). This does
not constitute a major problem in itself, for one could have just avoided
using this sort of a model. It follows, however, that  the solutions of the
models with linear scalar field and the inverse square potential discussed in
\cite{armendariz1, armendariz2} are isomorphic to those obtained in the
divergent models.
 
To see the relation between the models we work with the power law solutions. Consider a  typical model cosmology given by:

\be
\n{pl}
a=t^n,  \qquad V=\frac{\beta}{\phi^2},  \qquad   \phi=\phi_0\,t,
\ee
 
\no obtained by evaluating $F$ and $F_x$ at $x=x_0=-\phi_0^2$. We further use
$f=F(-\phi^2_0)$ and $f'=F_x(-\phi^2_0)$. Substituting these constants into
the Friedmann and $k$-field equations (\ref{00}), (\ref{kg}) we find that the
index $n$ and the slope of the potential $\beta$ are given by

\be
\n{sol}
n=\frac{1}{3}\frac{f+2\phi^2_0f'}{\phi^2_0f'}=\frac{2}{3\gamma}, \qquad
\beta=\frac{n}{f'},
\ee
 
On the other hand, if in the divergent model we choose the constants
$c=3n^2\phi^2_0/\beta$ and $b=-2n\phi_0/\beta$ we obtain the same solution.
Therefore all the power-law solutions obtained from the model with the linear
scalar field and the inverse square potential map into the solutions of the
divergent model with the same  potential.
 
Moreover, the following reasoning underlines the highly degenerate character
of the linear $k$-field solutions. Consider the series expansion of the
function  $F(x)$  around $x=x_0$. The background cosmology is  completely
determined by just the first two coefficients in the expansion $(f, f')$ and
the value of $\phi_0$ as seen from (\ref{sol}). Put in different words,  the
model is insensitive to keeping the first two coefficients in the expansion
of the function $F$ and the same value of $\phi_0$, but varying the rest of
the higher order terms. Since, given the same value of $\phi_0$, the first two
terms in the $F$ expansion  for the linear and divergent models coincide,
they should be thought of as equivalent. Therefore, the models with the
linear $k$-field and the inverse square potential possess a symmetry, or rather
a degeneracy in the sense that all the solutions for which the first two terms
of the expansion of the function $F$ around $x=x_0$ coincide are equivalent
among themselves and also equivalent to the divergent model. This  does not
happen with the solutions with the non-linearly behaved scalar field, and
suggests that physically the later are more acceptable, thus partially
removing the degeneracy of the solutions.
 
A more subtle distinction between the models, to completely remove the
degeneracy, would be probably seen by perturbing these solutions.  This,
however, is beyond the scope of the present paper.
 
%==================================================================
\section{Conclusions}
%==================================================================

In this paper we have studied particular solutions to the Einstein Equations
coupled to $k$-essence. Imposing  spatially flat  isotropic geometry we have
shown that different $k$-theory Lagrangians may lead to the same kinematical
evolution of the universe.

We have seen, however, that the linear $k$-field power-law solutions possess an
odd property of being isomorphic to a family of solutions with a divergent
speed of sound generated by the function $F^\infty$. This relation of
isomorphism induces, in fact, problems with the power-law solutions regardless
of the model (function$F$) as long as the potential is inverse square and the
field is linear, leading to consider different solutions. It has been recently
argued \cite{malquarti2} that the cosmological models based on linear $k$
field lead to serious problems. These problems are associated with the
behaviour of the models in the divergent sound speed region. We believe that
our findings relating the linear $k$-field models with the divergent models
sheds  new light on the reasons of the peculiar behavior of those models in
the region of the divergent speed of sound .

This work was supported by the University of Buenos Aires under Project X223.
 A.F. acknowledges the support of the University of the Basque Country
Grants 9/UPV00172.310-14456/2002, and The Spanish Science Ministry
Grant 1/CI-CYT 00172. 310-0018-12205/2000.

%\end{acknowledgments}

%==================================================================

\end{document}